\documentclass[12pt]{article}
\pdfoutput=1

\usepackage{amsmath}
\usepackage{amsfonts}
\usepackage{amssymb}
 
\usepackage[
      colorlinks=true,
      linkcolor=blue,
      urlcolor=blue,
      filecolor=black,
      citecolor=red,
      pdfstartview=FitV,
      pdftitle={},
        pdfauthor={Simon A. Gentle, Michael Gutperle},
        pdfsubject={},
        pdfkeywords={},
        pdfpagemode={},
        bookmarksopen=true
      ]{hyperref}

\usepackage{color}
\usepackage{graphicx}

\usepackage{sectsty}
\sectionfont{\large}

\renewcommand{\thefootnote}{\fnsymbol{footnote}}
\renewcommand{\thanks}[1]{\footnote{#1}}
\newcommand{\starttext}{
\setcounter{footnote}{0}
\renewcommand{\thefootnote}{\arabic{footnote}}}

\newcommand{\bea}{\begin{eqnarray}}
\newcommand{\eea}{\end{eqnarray}}
\newcommand{\ee}{\end{equation}}
\newcommand{\be}{\begin{equation}}
\newcommand{\<}{\langle}
\renewcommand{\>}{\rangle}

\def\Re{{\rm Re}}
\def\Im{{\rm Im}}

\DeclareMathOperator{\tr}{tr}
\usepackage{ dsfont } 
\def\id{\mathds{1}}
\renewcommand{\Im}{\operatorname{Im}}
\renewcommand{\Re}{\operatorname{Re}}

\long\def\symbolfootnote[#1]#2{\begingroup%
\def\thefootnote{\fnsymbol{footnote}}\footnote[#1]{#2}\endgroup}

\begin{document}
\setlength{\baselineskip}{18pt}

\starttext
\setcounter{footnote}{0}

\begin{flushright}
21$^{\mathrm{st}}$ July 2014
\end{flushright}

\bigskip

\begin{center}

{\Large \bf  Entanglement entropy of Wilson loops: \\[0.2cm] Holography  and matrix models}

\vskip 0.4in

{\large  Simon A.\ Gentle and Michael Gutperle}

\vskip .2in

{ \it Department of Physics and Astronomy }\\
{\it University of California, Los Angeles, CA 90095, USA}\\[0.5cm]
\href{mailto:sgentle@physics.ucla.edu}{\texttt{sgentle@physics.ucla.edu}}\texttt{, }\href{mailto:gutperle@physics.ucla.edu}{\texttt{gutperle@physics.ucla.edu}}

\end{center}

\begin{abstract}


A half-BPS circular Wilson loop in $\mathcal{N}=4$ $SU(N)$ supersymmetric Yang-Mills theory in an arbitrary representation is described by a Gaussian matrix model with a particular insertion.  
The additional entanglement entropy of a spherical region in the presence of such a  loop was recently  computed by Lewkowycz and Maldacena using exact matrix model results. In this note we utilize the supergravity solutions that are dual to such Wilson loops in a  representation with order $N^2$ boxes to calculate this entropy holographically.  Employing the matrix model results of  Gomis, Matsuura, Okuda and Trancanelli we express this holographic entanglement entropy in a form that can be compared with the calculation of Lewkowycz and Maldacena. We find complete agreement between the matrix model and holographic calculations.

\end{abstract}

\setcounter{equation}{0}
\setcounter{footnote}{0}

%
%
%
%
%
\newpage


\section{Introduction}
\setcounter{equation}{0}
\label{sec1}

In this note we investigate the additional entanglement entropy of a spherical region in ${\cal N}=4$ supersymmetric Yang-Mills theory in the presence of an insertion of a  half-BPS circular  Wilson loop  in a general representation of $SU(N)$ from two distinct points of view: from a matrix model and from gauge/gravity duality.

The expectation value  of such a  loop in the fundamental representation of $SU(N)$ was first computed in \cite{Erickson:2000af,Drukker:2000rr}. A special conformal transformation maps the circle to a straight line, whilst sending one point to infinity. It is known that the expectation value for the line is exactly unity and so the non-trivial value for the circular loop must come from the point at infinity. This was confirmed in \cite{Pestun:2007rz}, wherein it was shown using localization techniques that the circular loop is described by a Gaussian matrix model for arbitrary representations of $SU(N)$.

In \cite{Lewkowycz:2013laa}, Lewkowycz and Maldacena related this entanglement entropy to the expectation value of a circular loop and of the stress tensor in the presence of this loop by mapping the problem into the calculation of thermal entropy for a finite temperature field theory on a hyperbolic space \cite{Casini:2011kv}. Since both quantities can be calculated through localization by a matrix model, it is possible to obtain an expression for the entanglement entropy in the large $N$, large $\lambda$ limit  in an arbitrary representation.

The holographic description of  Wilson loops in Type IIB string theory  goes back to \cite{Maldacena:1998im,Rey:1998ik} wherein it was shown that  the Wilson loop   in the fundamental representation is described by a fundamental string in $AdS_5\times S^5$. For larger representations the fundamental string gets replaced by a probe D-brane.  It was shown in \cite{Gomis:2006sb,Gomis:2006im,Yamaguchi:2006tq}  that a  Wilson loop in  the $k$-th  symmetric  (or antisymmetric) representation is described by a D3 (or D5) brane with $k$ units of electric flux on its world volume. 

A general representation is characterized by a Young tableau.  If  the number of boxes  becomes of order $N^2$ then the probe-brane description breaks down and is replaced by  a fully back-reacted ``bubbling'' solution.  Such  solutions were first constructed in \cite{D'Hoker:2007fq}, building on the earlier work of \cite{Lunin:2006xr,D'Hoker:2007xy,D'Hoker:2007xz,Yamaguchi:2006te}.  Our first goal in this note is to calculate the entanglement entropy in the presence of a half-BPS circular  Wilson loop by applying the Ryu-Takayanagi prescription \cite{Ryu:2006bv,Ryu:2006ef}  to these static Type IIB supergravity solutions.

This holographic entanglement entropy can then be expressed in a form that makes  comparison with the matrix model calculation possible. We  show that in the saddle-point approximation of the matrix model, and at large $\lambda$, the two calculations agree. In our opinion this agreement   is  non-trivial since the two ways to calculate the entanglement entropy look very different from the outset. 
One can interpret this agreement as a non-trivial check of the calculation of  \cite{Lewkowycz:2013laa}, or alternatively as further confirmation of the map proposed  in  \cite{Okuda:2008px} between the supergravity solutions and the matrix model description of the circular Wilson loop.

Before moving forward, let us clarify the geometry of our setup. We are always interested in the circular Wilson loop. The entanglement entropy of the half-space that is intersected once by a circular loop  is conformally equivalent to a straight line threading a spherical region with the point at infinity included. We will find it more convenient to work with the latter setup when we compute the entanglement entropy holographically.

This note is organized as follows. In section~\ref{sec2} we review the matrix model description of our  Wilson loop and state the formula given by Lewkowycz and Maldacena for the entanglement entropy.  In section~\ref{sec3} we review the supergravity solutions dual to half-BPS Wilson loops  constructed in  \cite{D'Hoker:2007fq} and their relation to the matrix model data. In section~\ref{sec4} we calculate the entanglement entropy holographically and express it in a form that can be compared with the matrix model results. Careful attention is paid to the regularization of the resulting  integrals. In section~\ref{sec5} the matrix model and holographic calculations are compared and it is shown that if the matrix model saddle-point equations are satisfied then the two expressions agree. We close with a brief discussion of our results in~\ref{sec6}.
 Some calculational details regarding the regularization and holographic map of cut-offs are given in appendix~\ref{appa}. The proof of the equivalence between the matrix model  and holographic entanglement entropy is provided in appendix~\ref{appb}.
 
\section{Half-BPS Wilson loops and matrix models}
\setcounter{equation}{0}
\label{sec2}

The expectation value of a half-BPS circular Wilson loop in $\mathcal{N}=4$ $SU(N)$ supersymmetric Yang-Mills theory is described by a Gaussian matrix model.  This exact result was demonstrated in \cite{Pestun:2007rz} using localization techniques.  In particular,
\be \label{eq:generalW}
\< W_{\cal R} \> =\frac{1}{\cal Z}  \int [dM] \tr_{\cal R} e^{M'} \exp\left(-\frac{2 N}{\lambda} \tr M^2\right)
\ee
where $M$ is an $N\times N$ Hermitian matrix, $M' \equiv M - \frac{1}{N} ( \tr M) \id_{N\times N}$ is its trace-removed form and ${\cal R}$ is the representation of $SU(N)$.  In this note we focus on large representations for which the number of boxes in each row or column of the corresponding Young tableau is of order $N$.

One can evaluate $\<W_{\cal R}\>$ using saddle-point methods.  To leading order in the saddle-point approximation, i.e.\ at large $N$ with $\lambda$ held fixed, the normalized expectation value of the Wilson loop satisfies
\begin{equation}\label{eq:logW}
\log\, \< W_{\cal R} \>=-\left({\cal S}_{\mathrm{mat}}-{\cal S}_0\right)
\end{equation}
where ${\cal S}_{\mathrm{mat}}$ and ${\cal S}_0$ denote the on-shell effective action of the Gaussian matrix model with and without the insertion of the Wilson loop, respectively.  At large $\lambda$ it was shown in \cite{Okuda:2008px} that the effective action can be written as follows:
\begin{equation}\label{eq:Smat}
-\mathcal{S}_{\mathrm{mat}}= N\sum_{I=1}^{g+1}\int_{\mathcal{C}_I}dx\, \rho(x) \left( - \frac{2 N}{\lambda}\, x^2 + \hat{K}_I\, x \right) + N^2 \int_{\mathcal{C}\times\mathcal{C}}dx\, dy\, \rho(x)\, \rho(y) \log |x-y|
\end{equation}

Let us define the terms in this equation.  The matrix $M$ is decomposed into $g+1$ blocks of size $n_I\times n_I$.  For large $N$, the eigenvalues of  $M$ form a continuous distribution $\rho(x)$ over $g+1$ intervals $\mathcal{C}_I$.  The  $I^{\textrm{th}}$ interval contains a fraction $n_I/N$ of the eigenvalues and $\mathcal{C}$ is the union of these intervals.  The interactions between the eigenvalues on the intervals simplify at  large $\lambda$ to  the logarithmic repulsion term shown.  The Young tableau of interest consists of $g$ blocks and the $I^{\textrm{th}}$ block has  $n_I$ rows of length $K_I$.   We define $\hat{K}_I \equiv K_I - |\mathcal{R}|/N$, where $|{\cal R}|$ is the total number of boxes.\footnote{The parameters $K_I$ and $\hat K_I$  are  associated with  $U(N)$ and $SU(N)$ gauge groups, respectively \cite{Okuda:2008px}.}  We also note that $K_{g+1}=0$ and the following  relations:
\begin{equation}\label{eq:fillingfractions}
\frac{n_I}{N} = \int_{\mathcal{C}_I}dx\, \rho(x)\quad\textrm{and}\quad \sum_I\int_{\mathcal{C}_I}dx\, \rho(x)=\int_{\mathcal{C}}dx\, \rho(x)=1
\end{equation}

The eigenvalue distribution $\rho(x)$ satisfies the continuum version of the saddle-point equation:
\be\label{eq:SPE}
- 4 x + \frac{\lambda}{N}\,\hat{K}_I + 2 \mathrm{P}  \int_{\mathcal{C}} dy\,  \frac{\rho(y)}{x-y} =0 \quad \mathrm{for} \quad x\in {\cal C}_I
\ee
This is a set of singular integral equations that can be solved by introducing the resolvent $\omega(z)$, which takes the following form in the large $N$ limit:
\begin{equation}\label{eq:omega}
\omega(z) = \lambda \int_{\mathcal{C}} dx\, \frac{\rho(x)}{z-x}
\end{equation}
As a function of the spectral parameter $z$, this is  analytic  on the whole complex plane except on the intervals ${\cal C}_I$, where it has a discontinuity as one crosses each interval.  We can re-write \eqref{eq:SPE} in terms of these discontinuities as
\be\label{eq:SPE2}
- 4 x + \frac{\lambda}{N}\,\hat{K}_I + \omega_+(x)+\omega_-(x) =0 \quad \mathrm{for} \quad x\in {\cal C}_I
\ee
where $\omega_\pm(x)\equiv \omega(x\pm i \epsilon)$.

The action in the absence of the  loop is given by \eqref{eq:Smat} for a single interval (i.e.\ $g=0$), in which case the eigenvalues are distributed according to the Wigner semicircle rule:
\begin{align}
 \mathcal{S}_0 &= N^2\left( -\log\sqrt{\lambda}+\log 2 +3/4 \right) \label{eq:S0} \\
\rho_{(0)}(x) &= \frac{2}{\pi\lambda}\sqrt{\lambda-x^2} \quad\textrm{for}\quad x\in\left[-\sqrt{\lambda},\sqrt{\lambda}\right]  \label{eq:Wigner}
\end{align}
A useful result for a single interval  that we shall need later on is
\begin{equation}\label{eq:AdSlog}
\int_{\mathcal{C}\times\mathcal{C}}dx\, dy\, \rho(x)\, \rho(y) \log |x-y| =  \log\sqrt{\lambda}-\log 2 -\frac{1}{4} 
\end{equation}

Next we review the calculation of Lewkowycz and Maldacena \cite{Lewkowycz:2013laa}.    The quantity of interest is the   entanglement entropy relative to the vacuum of a spherical region of radius $R$ threaded by a half-BPS  circular Wilson loop.   They showed that this can be expressed as a sum of the  expectation value of this loop and the one-point function of the stress tensor in the presence of this loop.  The latter is fixed by conformal symmetry up to an overall coefficient $h_W$ known as the scaling weight  of the Wilson loop.  Their formula is 
\be\label{eq:EE}
\Delta S_{\cal A} = \log\, \<W_{\cal R}\> + 8\pi^2 h_W
\ee
using the sign convention of  \cite{Gomis:2008qa}.  It was shown in \cite{Gomis:2008qa} that the scaling weight is related to the difference  between the second moment of the matrix model eigenvalue distribution with the Wilson loop and without, $\Delta\rho_2\equiv\rho_2-\rho_2^{(0)}$,  via
\be\label{eq:hW}
h_W = -\frac{N^2}{3 \pi^2 \lambda}\, \Delta \rho_2
\ee
where
\be\label{eq:Deltarho2}
\rho_2 \equiv \int_{\cal C} dx\, \rho(x)\, x^2 \quad \mathrm{and} \quad \rho_2^{(0)} = \frac{\lambda}{4} 
\ee

In section~\ref{sec5} we will show that our holographic computation of the entanglement entropy  agrees precisely with \eqref{eq:EE}.

\section{Supergravity description of half-BPS Wilson loops}
\setcounter{equation}{0}
\label{sec3}

In this section we review the  features of the supergravity solutions that are important for the present work.  Their derivation and more details can be found in \cite{D'Hoker:2007fq}.  These static solutions have isometry group $SO(2,1)\times SO(3)\times SO(5)$ and preserve 16 out of the total 32 supersymmetries, which are the same symmetries as a  half-BPS  circular Wilson loop.    The ten-dimensional metric takes the form of a Janus-like ansatz \cite{Bak:2003jk} using a fibration of  $AdS_2\times S^2\times S^4$  over a two-dimensional Riemann surface $\Sigma$ with boundary $\partial\Sigma$. The metric   can be written in the form\footnote{We  deviate slightly from the notation in \cite{D'Hoker:2007fq}  and call a metric function $\sigma$ instead of $\rho$ to prevent confusion between the metric functions and the matrix model eigenvalue distribution.}
\begin{equation}\label{eq:generalmetric}
ds^2 = f_1^2\, ds^2_{AdS_2}+f^2_2\, ds^2_{S^2} +f_4^2\, ds^2_{S^4} + 4 \sigma^2 d\Sigma^2,\quad d\Sigma^2 = |dw|^2
\end{equation}
where $ds^2_{S^2}$ and $ds^2_{S^4}$ are the metrics on the unit radius two and four sphere, respectively. The metric on the unit radius Euclidean $AdS_2$ in Poincar\'{e}  half-plane coordinates is given by
\begin{equation}\label{eq:AdS2metric}
ds^2_{AdS_2} = \frac{dv^2+d\tau^2}{v^2}
\end{equation}

These half-BPS solutions are characterized by two harmonic functions $h_1,h_2$ defined on $\Sigma$. The metric functions are most easily expressed in terms of the following auxiliary quantities:
\begin{gather}
W= \partial_w h_1 \partial_{\bar w} h_2+\partial_w h_2 \partial_{\bar w} h_1, \quad \quad 
V=\partial_w h_1 \partial_{\bar w} h_2-\partial_w h_2 \partial_{\bar w} h_1\nonumber \\
N_1= 2 h_1 h_2\, \partial_w h_1 \partial_{\bar w} h_1 - h_1^2 W, \quad\quad
N_2= 2 h_1 h_2\, \partial_w h_2 \partial_{\bar w} h_2 - h_2^2 W \label{solpar1}
\end{gather}
The expressions for the dilaton $\Phi$ and the metric functions are then given by
\begin{gather}
e^{2\Phi}= -{N_2\over N_1}, \quad \quad \sigma^8= -{W^2 N_1 N_2\over h_1^4 h_2^4}\nonumber \\
f_1^4 = - 4 e^\Phi h_1^4 {W\over N_1}, \quad\quad f_2^4= 4 e^{-\Phi} h_2^4 {W\over N_2}, \quad\quad f_4^4 =4 e^{-\Phi} {N_2\over W}\label{solpar2}
\end{gather}
For the   regular solutions constructed in  \cite{D'Hoker:2007fq}   the Riemann surface $\Sigma$ is taken to be the lower half-plane  $w\in \mathbb{C}$, $\Im w<0$.  The $AdS_5\times S^5$ vacuum (i.e.\ no Wilson loop is present) is realized as follows:
\be
h_1 \sim \sqrt{1-w^2}+  \sqrt{1-\bar w^2}, \quad \quad  h_2\sim i(w- \bar w)
\ee
Note that in this case the harmonic function $h_1$ satisfies the following boundary conditions on the real line, which is the boundary of $\Sigma$: Neumann boundary conditions inside the interval  $\Re w\in [-1,1]$ and vanishing Dirichlet boundary conditions outside this interval. 

\begin{figure} [htb]
\centering
\includegraphics[scale=.5]{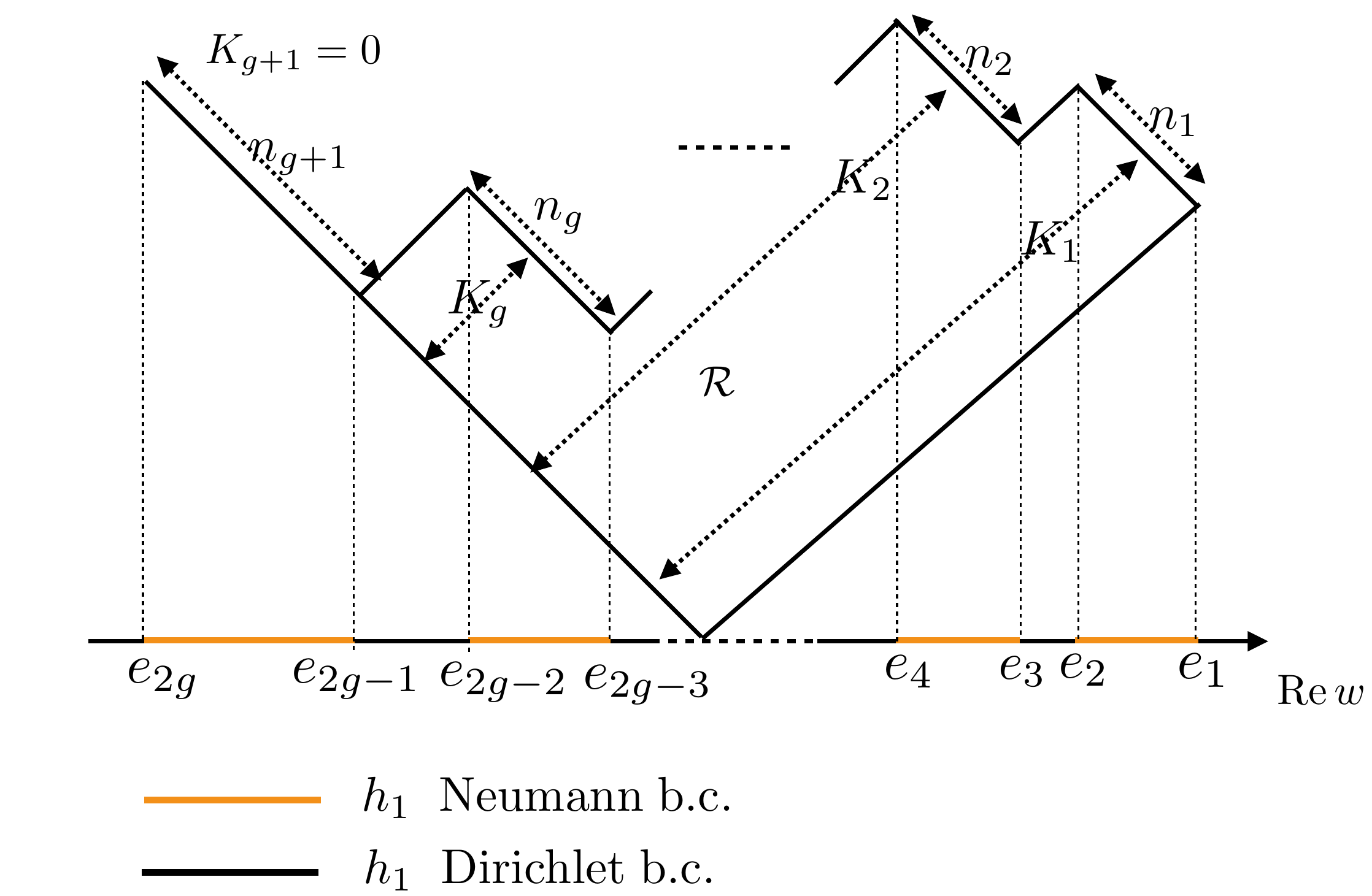}
\caption{The map  between   the supergravity data specified in (\ref{sugradata}) and the data of the  representation ${\cal R}$  of the circular Wilson loop specified by $\{n_I,K_I\}$.  The figure is adapted from  \cite{Okuda:2008px}. }
\label{fig1}
\end{figure}

 The  general regular solutions are constructed by modifying the boundary conditions for the harmonic function $h_1$.  A genus $g$ solution is characterized by $2g+2$  real numbers  
\be
 e_i\in \mathbb{R}, \quad i=1,2,\ldots, 2g;\quad  e_0=+\infty ,\quad e_{2g+1} = -\infty 
\ee
with the ordering $e_i> e_{i+1}$. The boundary conditions for $h_1$ alternate  as follows:
\be\label{sugradata}
\left. h_1\right|_{\Im w = 0} : \;\; 
\left\{
\begin{array}{ll}
 {\rm Neumann}, & \Re w \in [e_{2i} , e_{2i-1}]\\
 {\rm Dirichlet}, &  \Re w \in [e_{2i-1} , e_{2i-2}]   \\
\end{array}
\right.
\ee
For example, the explicit  $g=1$ solution  can be expressed in terms of elliptic integrals.  Full details of this solution, including formulae for the antisymmetric tensor fields, can be found in \cite{D'Hoker:2007fq,Okuda:2008px,Benichou:2011aa} but will not be needed in this paper.  It was shown in \cite{Okuda:2008px} that there is an exact identification between the data for the supergravity solution encoded in the boundary conditions (\ref{sugradata}) and the representation ${\cal R}$ of the circular Wilson loop --- see figure~\ref{fig1}.

A map between the supergravity solutions and the matrix model quantities was  also found  in \cite{Okuda:2008px}.  The harmonic functions are given in terms of the spectral parameter $z$ and matrix model  resolvent $\omega(z)$ via
\begin{equation}\label{eq:h1h2}
h_1 = \frac{i \alpha'}{8 g_s} \left[ 2(z-\bar{z}) -(\omega -\bar{\omega}) \right] \quad \textrm{and} \quad h_2 = \frac{i \alpha'}{4} \left(z-\bar{z}\right)
\end{equation}
Here we identify the spectral parameter $z$  with the coordinate we use on $\Sigma$: $z\equiv w$.  It takes values in the lower half-plane. In the following sections we will exploit this map to show that the holographic and matrix model calculations give the same results for the entanglement entropy of our Wilson loop.

\section{Holographic calculation of entanglement entropy}
\setcounter{equation}{0}
\label{sec4}

The Ryu-Takayanagi prescription \cite{Ryu:2006bv,Ryu:2006ef} states that the entanglement entropy of  a spatial region ${\cal A}$  is given by the area of a co-dimension two  minimal surface $\cal M$ in the bulk that is anchored on the $AdS$ boundary at $\partial {\cal A}$:
\be
S_{\cal A}= {A_{\mathrm{min}} \over 4 G_N^{(10)}}
\ee
Since we are dealing with static states of our CFT, this surface lies on a constant time slice. If this surface is not unique, we choose the one whose area is minimal among all such surfaces homologous to ${\cal A}$.\footnote{This minimal surface prescription  was recently established on a firm footing by the analysis of \cite{Lewkowycz:2013nqa}.}

The spacetime of interest is an $AdS_2\times S^2\times S^4$ fibration over $\Sigma$.  We consider a surface ${\cal M}$ parametrized by integrating over the $S^2$, $S^4$ and $\Sigma$ and choosing the spatial $AdS_2$ coordinate in (\ref{eq:AdS2metric}) to depend on $\Sigma$, i.e.\ $v=v(z,\bar z)$.  The area functional becomes
\be\label{minarea1}
A ({\cal M})= 2\, \textrm{Vol}(S^2)\, \textrm{Vol}(S^4) \int d^2z\, f_2^2\, f_4^4\, \sigma^2 \sqrt{1+\frac{f_1^2}{v^2 \sigma^2} \frac{\partial v}{\partial z} \frac{\partial v}{\partial \bar{z}}} 
\ee
Following  \cite{Jensen:2013lxa,Estes:2014hka}  it is easy to see that the minimal area surface is given by setting $v(z,\bar z)$ to a constant, since the second term under the square root in (\ref{minarea1}) is always positive and vanishes only for constant $v$. We will show in appendix~\ref{appa} that the choice $v=R$ in this $AdS_2$ slicing corresponds at the boundary to our desired region ${\cal A}$: a sphere of radius $R$  in Poincar\'{e} slicing.

The minimal area is therefore
\begin{align}
A_{\mathrm{min}}&=2\, \textrm{Vol}(S^2)\, \textrm{Vol}(S^4) \int d^2z\, f_2^2\, f_4^4\, \sigma^2 \nonumber\\
&=- {2^{9} \pi^3 \over 3}  \int d^2z\, \left\{2 h_2^2\, \partial_z h_1 \partial_{\bar z} h_1 - h_1 h_2 \left(\partial_z h_1 \partial_{\bar z}h_2 + \partial_{\bar z} h_1 \partial_{ z}h_2 \right)\right\}
\end{align}
where we used (\ref{solpar1}) and (\ref{solpar2})  to express $A_{\textrm{min}}$ in terms of the harmonic functions $h_{1,2}$.  

For the $g=1$ solution the entanglement entropy can in principle be evaluated by substituting the explicit expressions given in  \cite{D'Hoker:2007fq}  for  the harmonic functions  and performing the integrals. Since our goal  is to compare the holographic entanglement entropy to the matrix model calculation for arbitrary $g$, we instead use (\ref{eq:h1h2}) to rewrite the area of the minimal surface in terms of the matrix model resolvent $\omega(z)$:
\begin{align}
A_{\mathrm{min}}&=-\frac{  \pi^3 \alpha'^4}{6  g_s^2} \int d^2z\, \left\{  2 (z-\bar z)^2 ( \partial_z \omega +\partial_{\bar z} \bar\omega) -4(z-\bar z) (\omega-\bar \omega)\right. \nonumber \\
& \phantom{= -\frac{ \pi^3 \alpha'^4}{6  g_s^2} \int d^2z\, \Big\{  }  \left. - 2 (z-\bar z)^2 \partial_z \omega\, \partial_{\bar z}\bar \omega+  (z-\bar z) (\omega-\bar \omega) (\partial_z \omega+ \partial_{\bar z} \bar \omega)\right\}\label{amin3}
\end{align}
Note that we have dropped the $\partial_{ z}\bar \omega$ and $\partial_{\bar z} \omega$ terms from (\ref{amin3}): these are proportional to delta functions $\delta(z-x,\bar z-x)$, which integrate to zero against the $(z-\bar z)$ factors because $x$ in \eqref{eq:omega} is real. 

We rewrite the expression for $A_{\textrm{min}}$ by inserting the spectral representation \eqref{eq:omega} and performing the integration over $z$ after exchanging the order of integration. Since the integrals are divergent one has to take care with the regularization. The details of this calculation are presented in appendix~\ref{appa} and the final result for the holographic entanglement entropy is
\be
 S_{\cal A}=N^2\left[ \frac{R^2}{\varepsilon^2} -\log \frac{R}{\varepsilon}-\log \sqrt\lambda+\frac{3}{4} -\frac{2\rho_2}{3\lambda}  + \int_{\mathcal{C}\times\mathcal{C}}dx\, dy\, \rho(x)\, \rho(y) \log |x-y|   \right]  \label{eq:arearesult}
\ee
where $R$ is the radius of the spherical entangling region and $\varepsilon$ is the UV cut-off defined in the Fefferman-Graham chart near the $AdS$ boundary.

This is the result for a general number of intervals, describing a Wilson loop in a general representation ${\cal R}$.  The same expression for a single interval gives the area of the minimal surface in $AdS_5\times S^5$.  Thus, the result for the entanglement entropy of the vacuum is 
\begin{equation}\label{eq:AdSEE}
S_{\cal A}^{(0)}= N^2\left[ \frac{R^2}{\varepsilon^2} - \log \frac{2R}{\varepsilon} + \frac{1}{3} \right]
\end{equation}
where we used (\ref{eq:AdSlog}) and $\rho_2^{(0)}=\lambda/4$.  The  logarithmic term is universal and has coefficient $N^2$ as required.

The additional entanglement entropy due to the Wilson loop is found by subtracting the above two results:
\begin{equation}\label{eq:DeltaS}
\Delta S_{\cal A}=N^2\left[ \int_{\mathcal{C}\times\mathcal{C}}dx\, dy\, \rho(x)\, \rho(y) \log |x-y| - \frac{2\Delta\rho_2}{3\lambda}  -\left( \log\sqrt{\lambda}-\log 2 -\frac{1}{4}  \right) \right]
\end{equation}

\section{Comparison}
\setcounter{equation}{0}
\label{sec5}
Now we are ready to compare the holographic calculation with the matrix model result \eqref{eq:EE}.  Using \eqref{eq:logW}, \eqref{eq:Smat} and \eqref{eq:S0} we can write 
\begin{align}
\log\, \< W_{\cal R} \> &= N\sum_{I=1}^{g+1}\int_{\mathcal{C}_I}dx\, \rho(x) \left( - \frac{2 N}{\lambda}\, x^2 + \hat{K}_I\, x \right) + N^2 \int_{\mathcal{C}\times\mathcal{C}}dx\, dy\, \rho(x)\, \rho(y) \log |x-y| \nonumber \\
 &\phantom{=\ } +N^2\left( -\log\sqrt{\lambda}+\log 2 +\frac{3}{4} \right)
\end{align}
Adding this to the expression for the scaling weight $h_W$ in \eqref{eq:hW} we find that our result for  $\Delta S_{\cal A}$  in  \eqref{eq:DeltaS} appears, along with  two additional terms:
\be\label{eq:coneqa}
\log\, \< W_{\cal R} \> + 8\pi^2 h_W = \Delta S_{\cal A}  -{4N^2\over \lambda}\,  \Delta \rho_2 + { N} \sum_I \int_{C_I} dx\, \rho(x)\, \hat K_I\, x
\ee
In appendix~\ref{appb} we show that the last  two  terms on the right hand side of (\ref{eq:coneqa}) sum to zero, once we impose the saddle-point equation \eqref{eq:SPE2}.
Consequently we find complete agreement between the holographic calculation and the  Lewkowycz and 
Maldacena result.

\section{Discussion}
\setcounter{equation}{0}
\label{sec6}

In this  note we provided a proof of the agreement between two methods to calculate the entanglement 
entropy in the presence of a half-BPS circular Wilson loop: the replica method of Lewkowycz and 
Maldacena and  the (suitably-modified) holographic prescription of Ryu and Takayanagi. An essential 
ingredient in our  proof was the matrix model description of the expectation value of this  Wilson 
loop (and related moments) in the saddle-point approximation.

The original prescription for the calculation of  holographic entanglement entropy considered the area of minimal surfaces in $AdS$ spaces. Here we  generalized this prescription 
due to the fact that the spacetime is a fibration of $AdS_2\times S^2 \times S^4$ over a 
Riemann surface $\Sigma$. Specifically, our prescription takes the minimal surface to span the 
spheres as well as the Riemann surface $\Sigma$. Note that the same prescription has been 
used in related  holographic calculations of the boundary entropy of BPS interface solutions \cite{Chiodaroli:2010ur}, which are constructed using similar fibrations \cite{Chiodaroli:2009yw,Chiodaroli:2011nr}.  It was shown in \cite{Chiodaroli:2010ur} that the holographic boundary entropy  agreed with the CFT results \cite{Azeyanagi:2007qj}. 
In our opinion, the new  example of a highly non-trivial agreement found in the present note further strengthens the case that the generalized prescription is correct.

As mentioned in section~\ref{sec1}, we could equally well have choosen global coordinates  (i.e.\ the hyperbolic disk) in \eqref{eq:AdS2metric} and found the same minimal surface. The UV cut-off is blind to this difference because the coordinate transformation between Poincar\'{e} and global $AdS_2$ does not involve the five-dimensional radial coordinate. Consequently the result for $\Delta S_{\cal A}$  would not be modified.

Lewkowycz and Maldacena also calculated the entanglement entropy for the Wilson loop insertion in the three-dimensional  ${\cal N}=6$ supersymmetric Chern-Simons matter (ABJM) theory in  \cite{Lewkowycz:2013laa}. 
Unfortunately, we cannot conduct a similar consistency check for this case because the supergravity solutions analogous to the Wilson loop solution of \cite{D'Hoker:2007fq} are not known.  It would be  interesting to see if such solutions can be developed using the methods of  \cite{D'Hoker:2008wc}.

\section*{Acknowledgements}

It is a pleasure to thank  Matthew Headrick, Per Kraus, Aitor Lewkowycz and Mukund Rangamani for useful discussions. SAG would also like to thank the Aspen Center for Physics for hospitality during the concluding stages of this project.  This work was supported in part by National Science Foundation grants PHY-13-13986 and PHYS-1066293.

\newpage

\appendix 

\section{Integrals and  regularization}
\setcounter{equation}{0}
\label{appa}
In this appendix we carefully discuss the regularization and evaluation of the integrals that make up the area of the minimal surface (\ref{amin3}). For clarity we split the integrals into two terms and evaluate them separately:
\begin{align}
I_1&\equiv\int d^2z\, \left\{ 2 (z-\bar z)^2 ( \partial_z \omega +\partial_{\bar z} \bar\omega) -4(z-\bar z) (\omega-\bar \omega) \right\}\label{eq:I1}\\
I_2&\equiv  \int d^2z\, \left\{-2(z-\bar z)^2 \partial_z \omega\, \partial_{\bar z}\bar \omega + (z-\bar z) (\omega-\bar \omega) (\partial_z \omega+ \partial_{\bar z} \bar \omega)\right\} \label{eq:I2}
\end{align}
To evaluate these integrals we insert the spectral representation \eqref{eq:omega} for the resolvents $\omega$ and perform the integrals over $z$ first.  

First consider $I_1$, which is linear in $\rho$.  Working in the Cartesian coordinates $z=x+i\, y$ we obtain
\be
I_1=- 64\, \lambda \int_{\cal C}  dx_1\, \rho(x_1) \int_{-\infty}^\infty dx \int^0_{-\infty} dy\,   {y^4 \over ((x-x_1)^2+ y^2)^2}
\ee
This integral is quadratically divergent at large $y$.  Superficially it appears that one can remove $x_1$ from $ I_1$ by a shift in the integration variable. However, as is well known from the evaluation of Feynman diagrams, such arguments fail for integrals that have power law divergences.

To see this, we work in polar coordinates $z=\sqrt{\lambda}\, r\, e^{-i\phi}$ instead. The factor of $\sqrt{\lambda}$ will enable a cleaner identification of the Fefferman-Graham cut-off --- see the end of this appendix. We obtain
\be
I_1=- 64\, \lambda^4  \int_{\cal C}  dx_1\, \rho(x_1)\int_0^\infty  dr  \int_0^\pi d\phi\, {r^5 \sin^4 \phi\over (r^2 \lambda- 2 r x_1\sqrt{\lambda} \cos\phi+x_1^2)^2}
\ee
Note that the integral is quadratically divergent at large $r$.  To regularize this divergence we cut off the radial integration  at some large $r_c$. The angular integral can be performed and we find
\be
 \int_0^\pi d\phi\, {r^5 \sin^4 \phi\over (r^2 \lambda- 2 r x_1\sqrt{\lambda} \cos\phi+x_1^2)^2}=\left\{
\begin{array}{cc}
  {3\pi r \over 8\lambda^2}, &  r>\frac{|x_1|}{ \sqrt{\lambda}}    \\
  {3 \pi r^5 \over 8 x_1^4},& r<\frac{|x_1|}{ \sqrt{\lambda}} 
\end{array}
\right.
\ee
Performing the regulated integral over $r$ we obtain
\be \label{inta1}
I_1= - 64\, \lambda^4 \int_{\cal C}  dx_1\, \rho(x_1) \left[ \frac{3\pi}{8}\left(\int_{\frac{|x_1|}{ \sqrt{\lambda}}}^{r_c} dr\, \frac{r}{\lambda^2} + \int^{\frac{|x_1|}{ \sqrt{\lambda}}}_{0} dr\, \frac{r^5}{x_1^4}  \right)\right]=12\pi\lambda^2 \left(\frac{2\rho_2}{3\lambda}-r_c^2\right)
\ee
where we  used
\be
\int_{\cal C}  dx \, \rho(x) = 1 \quad\textrm{and}\quad \rho_2 = \int_{\cal C}  dx\, \rho(x)\, x^2
\ee
Note that in addition to the quadratically divergent piece, proportional to $r_c^2$, there is also a finite piece.

The  integral in  \eqref{eq:I2} is   quadratic in $\rho$ and can be expressed as
\begin{align}
I_2&= -\frac{\lambda^2}{2}\int_{\cal C}  dx_1\, \rho(x_1)\int_{\cal C}  dx_2\, \rho(x_2) \nonumber \\
&\phantom{=\ }\times \int d^2z\,   (z-\bar z)^4\, {(x_1-x_2)^2-(z-x_1) (\bar z-x_2)-(\bar z-x_1) (z-x_2)\over |z-x_1|^4 |z-x_2|^4}
\end{align}
where we have symmetrized appropriately. Note that the integral over $z$ is logarithmically divergent instead of quadratically divergent.  It is therefore possible to shift the integration variable as $z= x_1 + \sqrt{\lambda}\, r\, e^{-i\phi}$ such that the integral will only depend on $\Delta x=x_1-x_2$.  After this shift we find
\begin{align}
I_2 &=16\lambda^3 \int_{\cal C}  dx_1\, \rho(x_1)\int_{\cal C}  dx_2\, \rho(x_2)\nonumber \\ 
&\phantom{=\ }\times\int_0^{r_c}  dr  \int_0^\pi d\phi\, r\, {2r^2\lambda + 2 r\Delta x \sqrt{\lambda}  \cos\phi - \Delta x^2 \over (r^2 \lambda+ 2 r \Delta x\sqrt{\lambda} \cos\phi+\Delta x^2)^2}  \,  \sin^4\phi \label{int2til}
\end{align}
Using
\be
 \int_0^\pi d\phi\, r\, {2r^2\lambda + 2 r\Delta x \sqrt{\lambda}  \cos\phi - \Delta x^2\over (r^2 \lambda+ 2 r \Delta x\sqrt{\lambda} \cos\phi+\Delta x^2)^2}  \,  \sin^4\phi
=\left\{
\begin{array}{cc}
\frac{3\pi}{4   \lambda  r} - \frac{7\pi\Delta x^2}{8 \lambda ^2 r^3}, &  r>\frac{|\Delta x|}{ \sqrt{\lambda}}     \\
\frac{\pi  \lambda  r^3}{4 \Delta x^4}-\frac{3\pi r}{8\Delta x^2},  & r<\frac{|\Delta x|}{ \sqrt{\lambda}}
\end{array}
\right.
\ee
and dropping terms that tend to zero as $r_c\to\infty$, the regulated integral (\ref{int2til}) becomes
\be\label{inta2}
I_2 = 12 \pi \lambda^2 \left(\log r_c-\frac{3}{4}  +\log\sqrt{\lambda} -  \int_{{\cal C}\times {\cal C}}   dx_1\,dx_2\, \rho(x_1)\, \rho(x_2) \log |\Delta x| \right)
\ee

Next we substitute the results \eqref{inta1} and \eqref{inta2} into \eqref{amin3} in order to  evaluate the holographic entanglement entropy:
\begin{align}
S_{\cal A}&= \frac{A_{\mathrm{min}}}{4 G_N^{(10)}} \nonumber \\
 &=  \left(-\frac{  \pi^3 \alpha'^4}{6  g_s^2}\right) \frac{1}{2^5\pi^6\alpha'^4}\,(12 \pi) (4\pi g_s N)^2\left[ \frac{2\rho_2}{3\lambda} -r_c^2 +\log r_c-\frac{3}{4} + \log \sqrt\lambda \right. \nonumber\\
 &\phantom{=\ } \left.  -\int_{\mathcal{C}\times\mathcal{C}}dx\, dy\, \rho(x)\, \rho(y) \log |x-y|   \right] \nonumber\\
  &=N^2\left[ r_c^2 -\log r_c-\log \sqrt\lambda+\frac{3}{4} -\frac{2\rho_2}{3\lambda}  + \int_{\mathcal{C}\times\mathcal{C}}dx\, dy\, \rho(x)\, \rho(y) \log |x-y|   \right] \label{eq:almostfinalEE}
\end{align}
where we used $4 G_N^{(10)} = \frac{1}{4\pi} (2\pi)^7\alpha'^4$ and also $\lambda=4\pi g_s N$ where appropriate. 

We still need to show how the radial cut-off $r_c$ is related to the  UV cut-off.  At large $r$, any bubbling geometry of the form   \eqref{eq:generalmetric} asymptotes to $AdS_5\times S^5$:
\be\label{eq:asympAdS1}
ds^2 = L^2 \left\{ \frac{dr^2}{r^2} + r^2 \left( ds_{AdS_2}^2+ds_{S^2}^2 \right) +d\phi^2 + \sin^2\phi\, ds_{S^4}^2\right\}
\ee
with the $AdS_2$ metric given  in \eqref{eq:AdS2metric}.\footnote{To see this, substitute $z=\sqrt{\lambda}\, r\, e^{-i\phi}$ and  the $g=0$ resolvent into \eqref{eq:h1h2} and construct the metric.}  Any asymptotically $AdS$ metric may be written as a Fefferman-Graham expansion, at least locally, in the asymptotically $AdS$ region.  We write this as a power series in $u$ about $u=0$, which for us takes the form
\be\label{eq:asympAdS2}
ds^2 = L^2 \left\{ \frac{1}{u^2}\left(du^2 +d\tau^2 + dy^2 +y^2 ds_{S^2}^2\right)+d\chi^2+\sin^2\chi\, ds^2_{S^4}\right\}
\ee
plus subleading corrections.  Comparing \eqref{eq:asympAdS1} and \eqref{eq:asympAdS2}, at leading order we identify
\be\label{eq:chartmap}
\frac{u}{v} = \frac{1}{r}, \quad y = v,\quad \chi=\phi
\ee
Therefore, the large-$r$ cut-off is related to the UV cut-off $u=\varepsilon$ on the minimal surface $v=R$ near the boundary via
\be
r_c = \frac{R}{\varepsilon}
\ee
Substituting this result into \eqref{eq:almostfinalEE} we arrive at the final answer \eqref{eq:arearesult} for the entanglement entropy.

It is straightforward to show how the surface $v=R$ in $AdS_2$ slicing ends on a sphere of radius $R$ at the boundary in Poincar\'{e} slicing.  Near the boundary we have the map \eqref{eq:chartmap} between the two slicings.  It is well known (see \cite{Ryu:2006bv,Ryu:2006ef}, for example) that the equation for a minimal surface anchored on a boundary sphere of radius $R$ in Poincar\'{e} slicing \eqref{eq:asympAdS2} is 
\be
u(y)^2+y^2=R^2
\ee
Close to the boundary, the first term goes to zero and $y\to v$.  Thus we find $v=R$, as required.

\section{Proof of equivalence}
\setcounter{equation}{0}
\label{appb}
In this appendix we give the details of the proof that the matrix model and holographic entanglement entropies are equal. 
The relation we have to prove is 
\be\label{proofa}
-{4\over \lambda}\,  \Delta \rho_2 + {1\over N} \sum_I \int_{C_I} dx\, \rho(x)\, \hat K_I\, x =0
\ee
First we substitute for $\Delta\rho_2$ using \eqref{eq:Deltarho2} and deduce that the left-hand side of this relation can be written
\be\label{proofb}
\mathrm{LHS}= 1 + \sum_I \int_{C_I}  dx\, \rho(x)\,  x  \left( -{4\over \lambda}\, x+{1\over N}\, \hat K_I\right)
\ee
Next we impose the saddle-point equations \eqref{eq:SPE2} and find
\be\label{proofc}
\mathrm{LHS}=1-   \int_{-\infty}^{\infty}\,  dx\, \rho(x)\,  x\, {\omega_+(x) +\omega_- (x)\over \lambda}
\ee
We are able to extend the integration range to the real line since $\rho(x)$ vanishes outside the intervals ${\cal C}_I$.  Following the conventions of \cite{Okuda:2008px}, the eigenvalue density can be expressed in terms of the resolvents $\omega_\pm(x)= \omega(x\pm i\epsilon)$ as
\be
\rho(x)= {i\over 2\pi \lambda} \left( \omega_+(x)-\omega_-(x)\right)
\ee
and hence (\ref{proofc}) can be written as
\be
\mathrm{LHS}= 1 +{1\over 2 \pi i}\, {1\over \lambda^2}  \int_{-\infty}^{\infty}  dx\,  x \left(\omega^2_+(x)-\omega^2_-(x)\right)
\ee
Now we employ the integral representation of $\omega(z)$ given in \eqref{eq:omega} to find
\begin{align}
\mathrm{LHS}&= 1 +{1\over 2 \pi i}    \int_{-\infty}^{\infty} dx\, x \int_{\cal C} dx_1 \int_{\cal C} dx_2 \left( { \rho(x_1) \over x-x_1+i \epsilon} \, { \rho(x_2) \over x-x_2+i \epsilon}\right. \nonumber \\
& \phantom{= 1 +{1\over 2 \pi i}   \int_{-\infty}^{\infty} dx x \int_{\cal C} dx_1 \int_{\cal C} dx_2 \Big(\ } \left. -{ \rho(x_1) \over x-x_1-i \epsilon}\,  { \rho(x_2) \over x-x_2-i \epsilon}\right)
\end{align}
First we exchange the order of integration.  The relevant integral over $x$ can be performed using the residue theorem and we find
\be
 \int_{-\infty}^{\infty}dx\, x  \left( { 1 \over x-x_1+i \epsilon} \, { 1 \over x-x_2+i \epsilon} -{ 1 \over x-x_1-i \epsilon}\, {  1 \over x-x_2-i \epsilon}\right)=-2 \pi i
\ee
Then we are simply left with 
\be
\mathrm{LHS}=1- \int_{\cal C} dx_1 \rho(x_1) \int_{\cal C} dx_2 \rho(x_2)
\ee
which vanishes since the eigenvalue density is normalized to unity.

In conclusion, we have shown that
(\ref{proofa}) holds and thus the two expressions for the entanglement entropy are equal.



\end{document}